\documentclass[aps,twocolumn,showpacs,amsmath,amssymb,superscriptaddress]{revtex4-1}
\usepackage[utf8]{inputenc}

\usepackage{tikz}
\usepackage{graphicx}
\usepackage{amsfonts}
\usepackage{amsmath}
\usepackage{amssymb,bbold}
\usepackage{color}
\usepackage{bm}
\usepackage{bbm}
\usepackage{dsfont}
\usepackage{enumerate}
\usepackage{amsfonts, amsmath, amsthm, amssymb} 
\usepackage{mathtools}
\usepackage{array}
\usepackage[colorlinks,bookmarks=false,citecolor=blue,linkcolor=red,urlcolor=blue]{hyperref}

\begin{document}

\title{Reservoir-engineering shortcuts to adiabaticity}

\author{Rapha\"el Menu}
\affiliation{Theoretische  Physik,  Universit\"at  des  Saarlandes,  D-66123  Saarbr\"ucken,  Germany}
\author{Josias Langbehn}
\affiliation{Dahlem Center for Complex Quantum Systems and Fachbereich Physik,
Freie Universit\"at Berlin, Arnimallee 14, D-14195 Berlin, Germany}
\author{Christiane P. Koch}
\affiliation{Dahlem Center for Complex Quantum Systems and Fachbereich Physik,
Freie Universit\"at Berlin, Arnimallee 14, D-14195 Berlin, Germany}
\author{Giovanna Morigi}
\affiliation{Theoretische  Physik,  Universit\"at  des  Saarlandes,  D-66123  Saarbr\"ucken,  Germany}

\date{\today}

\begin{abstract}
We propose a protocol that achieves fast adiabatic transfer between two orthogonal states of a qubit by coupling with an ancilla. The qubit undergoes Landau-Zener dynamics, whereas the coupling realizes a time-dependent Hamiltonian, which is diagonal in the spin's instantaneous Landau-Zener eigenstates. The ancilla (or meter), in turn, couples to a thermal bath, such that the overall dynamics is incoherent. We analyse the protocol's fidelity as a function of the strength of the coupling and of the relaxation rate of the meter. When the meter's decay rate is the largest frequency scale of the dynamics, the spin dynamics is encompassed by a master equation describing dephasing of the spin in the instantaneous eigenbasis. In this regime the fidelity of adiabatic transfer improves as the bath temperature is increased. Surprisingly, the adiabatic transfer is significantly more efficient in the opposite regime, where the time scale of the ancilla dynamics is comparable to the characteristic spin time scale. Here, for low temperatures the coupling with the ancilla tends to suppress diabatic transitions via effective cooling. The protocol can be efficiently implemented by means of a pulsed, stroboscopic coupling with the ancilla and is robust against moderate fluctuations of the experimental parameters. 
\end{abstract}

\maketitle

\section{Introduction}

A powerful resource for quantum control is adiabatic dynamics because of its robustness against parameter fluctuations \cite{RevModPhys.90.015002, Shore}.  This comes at the cost of the operation time, which shall be sufficiently long in order to preserve adiabaticity \cite{de_grandi_adiabatic_2010}. In closed and finite systems, the lower bound to the time is set by the smallest frequency gap separating the target state from the excitations. In realistic settings, detrimental effects become increasingly important with time, such that they effectively determine a finite time window, the optimal processing time, in which the coherent adiabatic dynamics can be implemented. This implies a lower bound to the processing error \cite{Keck:2017, cepaite_counterdiabatic_2022}.  

A way to overcome this bottleneck is to develop protocols for implementing relatively fast and efficient adiabatic transformations. Strategies being discussed include the application of optimal control theory \cite{PhysRevA.85.033417} and the active use of projective measurements \cite{Kieferova:2014,PhysRevLett.108.080501}. 
The ultimate goal is to arbitrarily reduce the error of protocols based on quantum adiabatic dynamics. The formal equivalence of measurement and dissipative dynamics suggests the use of reservoir engineering. 

Here, we discuss a protocol for implementing fast adiabatic transfer by means of quantum reservoir engineering \cite{Poyatos:1996,Pielawa:2007,Kraus:2008,Morigi:2015,Roy:2020}. We start from the basic concept of quantum reservoir engineering, which tailors the coupling of a quantum system with a reservoir for efficient quantum state preparation, and we extend it with the objective of stabilizing the adiabatic quantum trajectory.  We consider a paradigmatic model of quantum adiabatic dynamics, the Landau-Zener Hamiltonian of a two-level system and design the coupling with an external bath. In our protocol the spin couples to an ancilla by means of a time dependent interaction or coupling Hamiltonian, that is diagonal in the instantaneous basis of the Landau-Zener Hamiltonian. Irreversible dynamics is introduced by means of a thermal bath, with which the ancilla thermalizes. The spin-ancilla coupling has the form of a quantum non-demolition (QND) type of Hamiltonian \cite{Braginsky:1980,haroche2013exploring,Yukalov2012}, where a measurement of the meter (ancilla) state projects the qubit onto an energy eigenstate. Differing from a QND measurement, however, the coupling is time-dependent and the coupling Hamiltonian is diagonal in the {\it instantaneous} Landau-Zener eigenbasis. We characterize the fidelity of adiabatic transfer as a function of the coupling strength with the ancilla, of the thermalization rate, and of the temperature of the external bath, and identify the regimes in which the coupling with the external bath leads to a significant gain in adiabaticity or in adiabatic transfer fidelity with respect to purely Hamiltonian dynamics. 

The paper is organized as follows. In Sec. \ref{Sec:LZ} we shortly review the properties of the adiabatic transfer probability of Landau-Zener Hamiltonian dynamics and then introduce our protocol based on quantum reservoir engineering. In Sec. \ref{Sec:NAME} we determine the fidelity of the adiabatic transfer taking into account the full quantum dynamics of the meter. We then investigate the role of non-adiabaticity and non-Markovianity of the spin-ancilla coupling on the fidelity of adiabatic transfer. In Sec. \ref{Sec:Exp} we analyse a stroboscopic, pulsed implementation of the protocol and simulate possible experimental imperfections, showing that the protocol is robust against moderate fluctuations of the experimental parameters. We conclude in Sec. \ref{Sec:Conclusions} and discuss perspectives of this work. The appendices provide details of the derivation and the benchmarking of the quantum adiabatic master equation for a weak quantum non demolition interaction.

\section{Landau-Zener Hamiltonian and QND measurement}
\label{Sec:LZ}

In this section we first review salient properties of the Landau-Zener dynamics. We then introduce the model at the center of our study, where the qubit, undergoing the Landau-Zener dynamics, is also coupled to a second quantum system, acting as an environment. The coupling Hamiltonian is time-dependent; at each instant of time it commutes with the Landau-Zener Hamiltonian and implements a dynamics which is reminiscent of quantum-non-demolition measurements. 

\subsection{Landau-Zener Hamiltonian}

The Landau-Zener model \cite{10011873546, doi:10.1098/rspa.1932.0165} is an example of exactly solvable dynamics and a workhorse of studies on adiabaticity \cite{de_grandi_adiabatic_2010,Dziarmaga:2010}. It describes the dynamics of a two-level system with the algebra of a spin $1/2$, whose Schr\"odinger equation is governed by the time-dependent Hamiltonian 
\begin{align}
\label{eq:H}
\hat{H}_\mathcal{S}(t) &= \dfrac{\epsilon t}{2}\hat\sigma_z + \dfrac{g}{2}\hat\sigma_x\,, 
 \notag\\
&= \dfrac{1}{2}\begin{pmatrix}
\epsilon t & g \\
g & -\epsilon t
\end{pmatrix},
\end{align}
where the matrix is reported in the eigenbasis of the Pauli matrix $\hat\sigma_z$, $| \uparrow \rangle\equiv (1,0)$, $|\downarrow \rangle\equiv(0,1)$. Here, the parameter $\epsilon$ is a positive constant that determines the sweeping rate of the Hamiltonian. As for $g$, it couples the energy levels and lifts the degeneracy observed at the crossing point $t=0$. For convenience we have set $\hbar=1$. 

\begin{figure}
\centering
\includegraphics[width = \columnwidth]{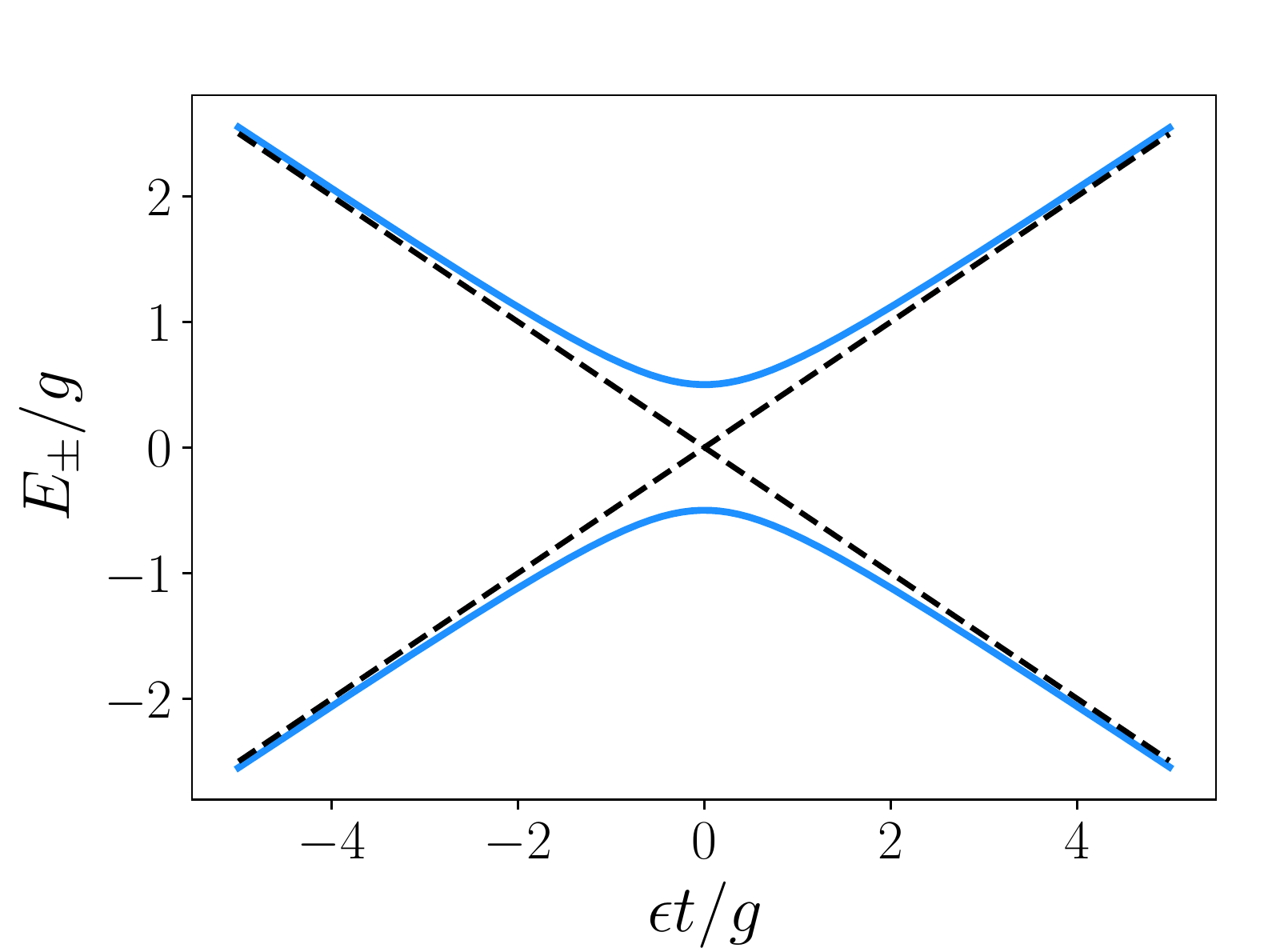}
\caption{Instantaneous eigenenergies of the Landau-Zener Hamiltonian \eqref{eq:H} as a function of time (in units of $g/\epsilon$). The blue curves correspond to the energies in units of $g$, which determines to the gap at the anticrossing $t=0$. The dashed lines correspond to the eigenenergies for $g=0$.}
\label{Fig1}
\end{figure}

Hamiltonian \eqref{eq:H} is diagonal in the so-called adiabatic eigenbasis $\vert \pm \rangle_t$, which is connected to the $\sigma_z$-eigenbasis by the relations
\begin{eqnarray*}
\vert + \rangle_t &=&+\cos(\theta(t)/2) \vert \uparrow \rangle+ \sin(\theta(t)/2)\vert \downarrow \rangle\,,\\
\vert - \rangle_t &=&-\sin(\theta(t)/2)\vert \uparrow \rangle+ \cos(\theta(t)/2)\vert \downarrow \rangle\,,
\end{eqnarray*}
with $\tan \theta(t) = g/\epsilon t$ \cite{damski_simplest_2005}.  The corresponding instantaneous eigenenergies $E_{\pm}(t) = \pm \sqrt{ g^2 + \epsilon^2 t^2 }/2$ are displayed in Fig.~\ref{Fig1} as a function of $t$. 
The lower branch is associated with state $\vert - \rangle_t$, corresponding to $\vert \uparrow \rangle$ at $t_1 \to -\infty$ and to $\vert \downarrow \rangle$ at $t_2\to +\infty$. The opposite holds for the state $\vert + \rangle_t$ in the upper branch.

The dynamics of the Landau-Zener model is governed by a time-dependent Schr\"odinger equation, which admits an analytical solution in terms of parabolic cylinder functions \cite{abramowitz+stegun}. A quantity relevant to our study is the probability $P(t_1,t_2)$ of a diabatic transfer from state $\vert - \rangle_{t_1}$ to state $\vert + \rangle_{t_2}$:
\begin{equation}
\label{eq:T}
P(t_1,t_2)=\vert _{t_1}\langle + \vert \hat{U}(t_1,t_2)\vert - \rangle_{t_2}  \vert^2 \,,
\end{equation}
where $\hat{U}(t_1,t_2)$ is the evolution operator solving the Schr\"odinger equation with the time-dependent Hamiltonian \eqref{eq:H}. The probability $P$ quantifies the deviation from adiabaticity and thus provides the error, or the infidelity, of the operation. We define the infidelity $T$ as the limit of $P$ when $t_1\to-\infty$ and $t_2\to +\infty$. In this limit it takes the well-known form \cite{10011873546,doi:10.1098/rspa.1932.0165}
\begin{equation}
T \equiv\lim_{\substack{t_1 \to -\infty \\ t_2 \to +\infty}}P(t_1,t_2)= \exp\left(-\pi g^2/2\epsilon\right)\label{Eq_3}\,,
\end{equation}
which shows that the infidelity is controlled by the ratio between the energy gap $g$ at the avoided crossing point and the sweeping rate $\epsilon$. One typically distinguishes two extremal behaviours: (i) For $g \gg \sqrt{\epsilon}$, the gap is sufficiently wide to penalize the transition, enabling the system to evolve smoothly from $\vert - \rangle_{-\infty} = \vert \uparrow \rangle$ to $\vert - \rangle_{+\infty} = \vert \downarrow \rangle$. (ii) On the other hand, if $g \ll \sqrt{\epsilon}$, the system does not have time to adjust to the change of parameters and the state gets promoted from the low-energy branch to the high-energy one.

For a finite time window, the transition probability is dominated by the term \cite{de_grandi_adiabatic_2010}
\begin{equation}
    P(t_1,t_2)\simeq \dfrac{\epsilon^2}{16g^4}\left( \dfrac{g^6}{(g^2 + \epsilon^2 t_1^2)^3} + \dfrac{g^6}{(g^2 + \epsilon^2 t_2^2)^3}\right)\label{Eq:finite}\,,
\end{equation}
and thus scales algebraically with $\epsilon/g^2$. Equation \eqref{Eq:finite} is reported to be valid for sufficiently long times $t_2$, for which the oscillations are damped out, and neglects higher-order corrections. We note that the latter also include the terms determining the asymptotic behaviour of Eq. \eqref{Eq_3} for $t_2,\vert t_1\vert \to \infty$. 

\subsection{Quantum non-demolition measurement in the adiabatic basis}
\label{Sec_4}

Projective measurements are an essential ingredient of quantum mechanics. They are physically implemented by coupling the system of interest to a second physical system, acting as a meter for the quantity to measure \cite{braginsky_khalili_thorne_1992}. As a result, the effective dynamics of the system is generally incoherent, since the measurement destroys quantum superpositions between eigenstates at different eigenvalues. 
At the same time, the incoherent dynamics due to the meter can suppress transitions to states outside the target Hilbert space, realizing an effective quantum Zeno dynamics \cite{PhysRevA.86.032120}. 

In this work we propose to design the coupling to the meter such that it suppresses diabatic transitions by performing a measurement in the instantaneous Landau-Zener eigenbasis in the spirit of quantum reservoir engineering \cite{Poyatos:1996,Pielawa:2007,Kraus:2008,Morigi:2015,Roy:2020}. This is done by coupling the spin with an ancilla by means of  the Hamiltonian $\hat{H}_\mathcal{SM}(t)$ defined in the Hilbert space of qubit and ancilla (which we here and below also denote by "meter"),
\begin{equation}
\hat{H}_\mathcal{SM}(t)= \hat{H}_\mathcal{S}(t)+\hat{H}_\mathcal{M}+ \hat{H}_{QND}(t)\,,
\label{H:tot}
\end{equation}
where  $\hat{H}_\mathcal{M}$ is the meter's Hamiltonian in the absence of the coupling and $ \hat{H}_{QND}(t)$ is the qubit-meter coupling. The latter is time dependent and takes the form
\begin{equation}
    \hat{H}_{QND}(t) = \hat{H}_\mathcal{S}(t) \otimes \hat{X}_\mathcal{M} \,,\label{Eq:QND}
\end{equation}
with $\hat{H}_\mathcal{S}(t)$ given in Eq. \eqref{eq:H} and $\hat X_M$ an operator acting on the Hilbert space of the meter (ancilla). 
Hamiltonian $\hat{H}_{QND}$ commutes with $\hat{H}_{\mathcal{S}}$ at each instant of time. For $\epsilon=0$ (no time dependence) 
it realizes a quantum non-demolition measurement: there is no energy exchange between meter and qubit and the measurement of the meter (ancilla) allows one to measure the energy of the qubit with arbitrary precision \cite{Braginsky:1980,haroche2013exploring,Yukalov2012}. 

QND-type of dynamics have been realized in several setups. In most cases, system and meter are qubit and harmonic oscillator with interchangeable roles. In microwave cavity QED, for instance, the system is a high-finesse mode of the cavity, the meter a Rydberg transition of atoms flying through the cavity \cite{RevModPhys.85.1083}. QND is at the basis of spin squeezing protocols using the mode of a resonator as a meter \cite{Vuletic:2017}. Most recently, a QND based protocol has been proposed for determining the spectrum of a spin chain by using the common vibrational mode as meter \cite{yang_quantum_2020}.  With respect to these examples, the peculiarity of Eq. \eqref{Eq:QND} is that $\hat{H}_{QND}(t)$ is time dependent and specifically diagonal in the adiabatic eigenbasis. 

We note that here the analogy with the measurement requires that the ancilla state is instantaneously reset after the interaction \cite{Englert:2003,Pielawa:2007,Morigi:2015,Roy:2020}, such that the correlations between ancilla and system generated by the interactions are destroyed by projecting the ancilla onto the initial state. In our dynamics this occurs on a finite time scale and is realised by the coupling of the ancilla with a thermal bath, with which the ancilla equilibrates. We will show that this time scale is indeed relevant and that the protocol works most efficiently when retardation effects are important. 

In the rest of this paper we discuss the qubit's effective dynamics generated by the QND coupling of Hamiltonian $\hat H_{QND}$, Eq.\ \eqref{Eq:QND}, assuming that the meter is a damped harmonic oscillator. 

\section{Case study: qubit coupled to a cavity ancilla}
\label{Sec:NAME}

We now analyse the dynamics governed by the meter-qubit Hamiltonian with the QND coupling as in Eq. \eqref{Eq:QND}. We specifically investigate whether and under which conditions the QND coupling suppresses diabatic transitions. In what follows, we assume that the meter (ancilla) is a damped oscillator. We describe the meter's damping by means of a Liouvillian and   numerically determine the infidelity of the adiabatic transfer as a function of the parameters characterizing the ancilla's dynamics. 

\subsection{Continuous QND coupling to a damped oscillator}

\begin{figure*}
\centering
\includegraphics[width = \textwidth]{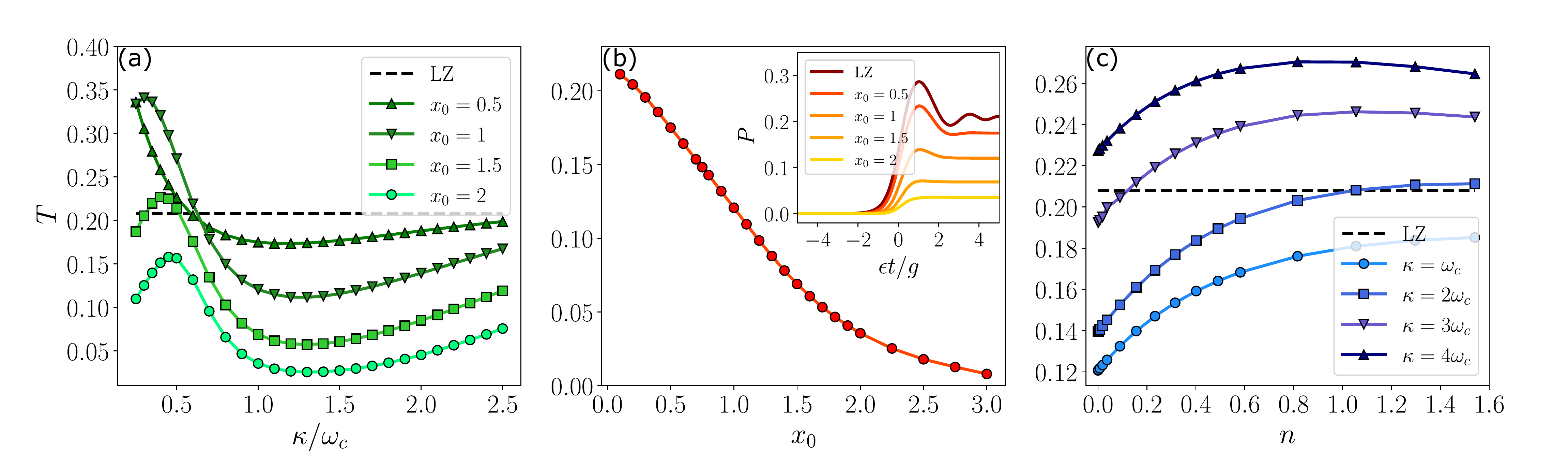}
\caption{Infidelity $T$ as a function of (a) the damping rate $\kappa$, computed for fixed parameters $\omega_c = g$, and $\beta = 10/\omega_c$ ($n\sim 4\times 10^{-5}$), (b) the qubit-meter coupling amplitude $x_0$, for $\omega_c = g$, $\kappa = \omega_c$ and $\beta=10/\omega_c$, and (c) the average occupancy $n=1/(\exp(\beta\omega_c)-1)$ of the boson bath, for $\omega_c = g$ and $x_0 = 1$. In all panels the adiabaticity parameter is $g^2/\epsilon = 1$, i.e. non-adiabatic effects in the bare dynamics of the qubit are expected to be sizeable. The inset of subplot (b) displays the evolution of the transfer probability $P$ for some values of $x_0$. All results are obtained by numerically solving Eq.~\eqref{Lindblad} for an oscillator of maximum occupancy $n_\mathrm{max} = 50$.}
\label{Fig:6}
\end{figure*}

Let now $H_\mathcal{M}=\omega_c \hat{a}^\dagger \hat{a}$ be the Hamiltonian of a harmonic oscillator, where $\hat{a}^\dagger$ and $\hat{a}$ create and annihilate, respectively, a quantum of energy $\omega_c$. The operator $ \hat{X}_\mathcal M$ of the QND Hamiltonian, Eq. \eqref{Eq:QND}, is taken to be 
$$ \hat{X}_\mathcal M= x_0  (\hat{a} + \hat{a}^\dagger)\,,$$
where $x_0$ is a dimensionless parameter scaling the QND coupling. The oscillator, moreover, couples to a thermal bath, which we assume to be Markovian. The resulting dynamics is governed by the Lindblad equation
\begin{align}
    \dfrac{\partial}{\partial t}\hat{\rho} =& -i[\hat{H}_{\mathcal{SM}}(t), \hat{\rho}] + \kappa(n+1)\left( \hat{a} \hat{\rho} \hat{a}^\dagger - \dfrac{1}{2}\lbrace \hat{a}^\dagger \hat{a}, \hat{\rho} \rbrace\right)\notag\\
    &+\kappa n \left( \hat{a}^\dagger \hat{\rho} \hat{a} - \dfrac{1}{2}\lbrace \hat{a} \hat{a}^\dagger, \hat{\rho} \rbrace\right)\label{Lindblad}\,,
\end{align}
where 
$\kappa$ is the oscillator's damping and  $$n=1/(\exp(\beta\omega_c)-1)\,$$
is the mean occupancy of the cavity imposed by the bath with inverse temperature $\beta$.

Using Eq. \eqref{Lindblad} we numerically determine the transfer probability integrating over a finite time interval $[-t,t]$. Exemplary dynamics of the transition probability as a function of time are depicted in the inset of Fig. \ref{Fig:6}(b): In general, the coupling to the meter tends to damp the typical oscillations of the Landau-Zener dynamics and leads to a faster relaxation towards the asymptotic value. 
In Fig.\ \ref{Fig:6} the infidelity is analysed as a function of the damping rate $\kappa$, of the coupling amplitude $x_0$, and of the average occupancy $n$. We start with the dependence on $\kappa$ and first note that in the limit $\kappa\to\infty$ the oscillator instantaneously relaxes to the thermal state. For large but finite $\kappa$ the oscillator is in a coherent state, whose amplitude is determined by the Landau-Zener Hamiltonian: This is the limit of a projective measurement, the state of the qubit is inferred by measuring the oscillator amplitude. Figure \ref{Fig:6}(a) displays the infidelity as a function of the damping rate $\kappa$ and for different values of $x_0$. For small $\kappa$, the infidelity decreases down to a local minimum that lies below the Landau-Zener prediction. In the limit $\kappa\gg x_0,\epsilon/g^2$, instead, the infidelity tends to increase. This limit will be extensively discussed in Sec. \ref{Sec:AME}.  

Increasing $x_0$ at fixed $\kappa$, on the other hand, is analogous to increasing the importance of retardation (or memory) effects on the dynamics. In Fig. \ref{Fig:6}(b) the infidelity is displayed as a function of $x_0$: for increasing values of $x_0$, the infidelity decreases down to a non-vanishing asymptotic value. This behavior is due to transitions between the instantaneous eigenstates. At large $x_0$ (and vanishing temperatures) these transitions mainly consist of decay from the upper to the lower energy branch, thus cooling the qubit into the instantaneous lower eigenstate and thereby correcting for the unwanted diabatic transitions. This behavior can substantially change when increasing the temperature (correspondingly, increasing $n$), as visible in Fig. \ref{Fig:6}(c). Here, the initial monotonous increase of the infidelity as a function of the thermal occupation $n$ is due to an increasing number of diabatic transitions from the lower to the upper branch. Remarkably, this tendency is reverted for large values of $\kappa$. We will show that in this limit the coupling with the ancilla leads to an effective dephasing in the instantaneous Landau-Zener eigenbasis, which tends to preserve the adiabatic dynamics and whose rate is proportional to the thermal occupation $n$. We here remark that similar features have been discussed in Refs.  \cite{PhysRevB.96.054301, PhysRevLett.97.200404} for a different implementation of the spin master equation. 
 
 \subsection{Memory effects and instantaneous gap}
 
 The analysis of the infidelity of our protocol exhibits a rich variety of behaviours and regimes. In some cases we observe an improvement of the adiabaticity compared to the Landau-Zener prediction. Particularly striking is that increasing qubit-cavity coupling leads to decreasing the infidelity, suggesting that memory effects could help the adiabatic transfer. In order to better understand whether there is a connection, we analyse non-Markovianity in the qubit's dynamics. Non-Markovianity in the evolution of a density matrix can be probed by several type of measures \cite{RevModPhys.88.021002}. In the following, we quantify non-Markovianity in the qubit by using the measure based on information backflow to the system \cite{PhysRevA.81.062115}, later on referred to as Non-Markovianity (NM) measure. A non-vanishing NM measure signals non-Markovian evolution and the presence of memory effects in the dynamics of the system.
 
 The measure $\mathcal{N}$ of Ref. \cite{PhysRevA.81.062115}  quantifies the revival of distinguishability between two states as a function of time. While for Markovian dynamics the distinguishability of any two states follows a monotonous decay, non-Markovian dynamics are characterized by a temporarily increasing distinguishability. The measure $\mathcal{N}$ of a dynamical map $\Phi$ determines distinguishability via the trace distance and considers the pair of initial states $\rho_{1,2}(0)$ that exhibits the maximum increase in distinguishability, 
\begin{align}
\sigma\left(t, \rho_{1,2}(0) \right) &= \frac{d}{dt}\mathrm{tr}\left |\rho_1(t) - \rho_2(t) \right | \\
\mathcal{N}\left( \Phi \right) &= \max_{\rho_{1,2}(0)} \int_{\sigma>0} dt\, \sigma\left(t, \rho_{1,2}(0) \right)\,,
\end{align}

\begin{figure}
    \centering
    \includegraphics[width = \columnwidth]{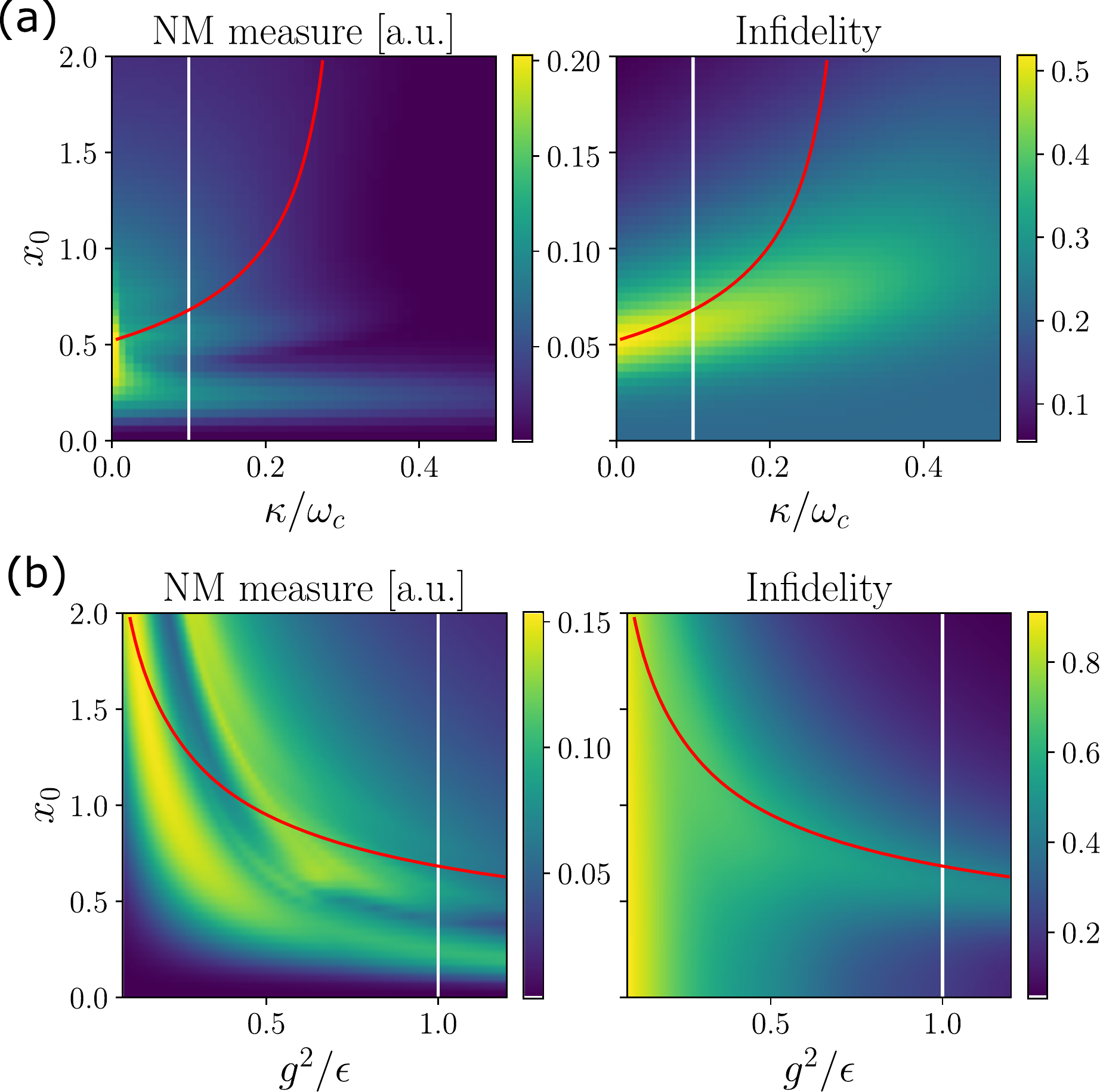}
    \caption{ (color online) NM measure (left panels) and value of the infidelity $T$ (right panels)  (a) as a function of the damping rate of the meter $\kappa/\omega_c$ and the QND coupling $x_0$ computed for the adiabatic parameter $g^2/\epsilon=1$ and (b) as a function of the adiabatic parameter $g^2/\epsilon$ and the QND coupling $x_0$ computed at a fixed damping rate $\kappa/\omega_c=0.1$. On all plots, the red line stands for the minimum of the effective real gap $\Delta_R$, see text for details, while the vertical white line represents the intersection between the considered $(\kappa/\omega_c, x_0)$ and $(g^2/\epsilon, x_0)$ planes. All calculations are performed for $\beta =10/\omega_c$ ($n\sim 4\times 10^{-5}$).}
    \label{fig:5}
\end{figure}

Figure~\ref{fig:5} displays the NM measure $\mathcal{N}$ and the corresponding infidelity for representative parameter regimes. As one would expect, non-Markovian effects are sizeable for weak damping as the meter is not reset fast enough. Yet, the NM measure exhibits one or more maxima as a function of the meter-qubit coupling $x_0$. 
In none of the two cases considered, we can identify a clear correlation between the behaviour of the NM measure and the infidelity. We conclude that there is no simple relationship between memory effects and adiabaticity.
  
 In addition, we also inspect the instantaneous gap between the two instantaneous eigenstates of the qubit at the crossing point:
$$\Delta_R = \Delta E(0)(1 + 2x_0 \langle \hat{a} + \hat{a}^\dagger \rangle)_{t=0}\,,$$
which we simply obtain by tracing out the oscillator. The red lines in Fig. \ref{fig:5} show the parameter values where $\Delta_R$ is minimum. This curve has a good overlap with the maximum of the infidelity on a large parameter interval, suggesting that high-fidelity adiabatic transfer can be achieved by designing an effective gap by means of an external environment.

\subsection{QND measurement in the Markovian regime}
\label{Sec:AME}

The purpose of this section is to discuss the equation governing the effective dynamics of the qubit when the oscillator's variable can be eliminated from the qubit's equation of motion. This occurs in the overdamped limit, when the meter quickly relaxes to a steady state. In this regime, the coupling with the meter induces an incoherent dynamics of the qubit. The resulting master equation describes an effective dephasing in the adiabatic basis, whose net effect is to enforce adiabatic transfer. In this case the infidelity can be analytically determined in some limits. We recall that the dynamics for master equations with QND coupling between system and reservoir have been discussed, for instance, in Ref. \cite{Gordon:2010}. Differing from those works, we emphasize that in our case the coupling is time-dependent and proportional to the Landau-Zener Hamiltonian. 

In what follows, we sketch  the derivation of the master equation following the steps of Ref. \cite{Albash_2015} and for a generic system acting as a meter. We then analyse the predictions of the master equation.

\subsubsection{Basic assumptions}

Our starting point is the von-Neumann equation for meter and qubit with Hamiltonian \eqref{H:tot}. The meter is assumed to be in a thermal state at inverse temperature $\beta$.  The construction of the quantum adiabatic master equation requires assumptions on the different energy and time scales. We start with the minimal gap $g$ between the two energy branches: In order to ascertain the adiabaticity of the coherent part of the dynamics for the LZ system, we shall take the limit where $\epsilon/g^2 \ll 1$. We also require that  the meter relaxes at time scales $\tau_M$ over which one can consider the Hamiltonian $\hat{H}_\mathcal{S}(t)$ to be constant, resulting in $\epsilon\tau_M^2\ll 1$. For example, when the meter is an overdamped oscillator, then $\tau_M\sim 1/\kappa$ when $\kappa$ is the largest rate of Eq. \eqref{Lindblad}, and the second condition corresponds to the inequality  $\epsilon\ll \kappa^2$. Under these conditions, the incoherent part of the dynamics is predominantly adiabatic since the first non-adiabatic corrections scale like $\epsilon/g^2$ and are thus of higher order. 

We follow the procedure outlined in Ref. \cite{Albash_2015}  and obtain the master equation for the qubit density matrix $\hat{\rho}_S = \mathrm{Tr}_\mathcal{M}[\hat{\rho}]$:
 \begin{widetext}
\begin{align}
    \partial_t{\tilde{\rho}_\mathcal{S}}(t) = - \int^{+\infty}_0{\mathrm{d}\tau}& \left\lbrace \tilde{H}_\mathcal{S} (t) \tilde{H}_\mathcal{S} (t - \tau)\tilde{\rho}_\mathcal{S}(t)-\tilde{H}_\mathcal{S} (t - \tau) \tilde{\rho}_\mathcal{S}(t) \tilde{H}_\mathcal{S} (t)\right\rbrace\mathcal{C}_{XX}(\tau,0)\notag\\
    -\int^{+\infty}_0{\mathrm{d}\tau}& \left\lbrace \tilde{\rho}_\mathcal{S}(t)\tilde{H}_\mathcal{S} (t-\tau) \tilde{H}_\mathcal{S} (t)-\tilde{H}_\mathcal{S} (t) \tilde{\rho}_\mathcal{S}(t) \tilde{H}_\mathcal{S} (t - \tau)\right\rbrace\mathcal{C}_{XX}(0, \tau)\label{Eq:MEq}\,,
\end{align}
\end{widetext} 
where the details of the derivation are provided in the Appendix \ref{App_A}. Equation \eqref{Eq:MEq} is reported in the interaction picture, where the operators $\tilde{A} = \tilde{H}_\mathcal{S}$ and $\tilde{\rho}$ are related to the operators in the laboratory frame by 
\begin{equation}
 \tilde{A}(t) = \hat U_\mathcal{S}^\dagger(t,0)\hat A(t)\hat U_\mathcal{S}(t,0),
\end{equation}
and $\hat U_\mathcal{S}(t,0)$ is the evolution operator of the qubit's time-dependent Schr\"odinger equation:
\begin{align}
\label{eq:U}
    \hat{U}_\mathcal{S}(t,0) &= \mathcal{T}\exp\left[-i\int_{0}^t{\mathrm{d}t'\hat{H}_{\mathcal{S}}(t')}\right]\,,
\end{align}
with $\mathcal T$ the symbol for time ordering. The scalar function $\mathcal{C}_{XX}(t, t')$ is the auto-correlation function of the observable $\hat{X}_\mathcal{M}$,
\begin{eqnarray}
\mathcal{C}_{XX}(t, t')&=&\langle \tilde{X}_\mathcal{M}(t) \tilde{X}_\mathcal{M}(t') \rangle\\
&=&{\rm Tr}\{{\rm e}^{{\rm i}\hat{H}_\mathcal{M}t}\hat{X}_\mathcal{M}{\rm e}^{-{\rm i}\hat{H}_\mathcal{M}(t-t')}\hat{X}_\mathcal{M}{\rm e}^{-{\rm i}\hat{H}_\mathcal{M}t'}\hat{\rho}_\mathcal{M}\}\,,\nonumber
\end{eqnarray}
with $\hat{\rho}_\mathcal{M}$ the density matrix of the meter. Since the meter's state is assumed stationary over the time scale of the qubit evolution, $\mathcal{C}_{XX}(t, t')=\mathcal{C}_{XX}(t- t',0)=\mathcal{C}_{XX}(0,t'-t)$. 

We expand the evolution operator $\hat{U}_\mathcal{S}$ to first order in the parameter $\epsilon/g^2$ in order to evaluate the integrals in Eq. \eqref{Eq:MEq}. We approximate Eq. \eqref{eq:U} by the expression
\begin{equation}
    \hat U_\mathcal{S}(t,t')=\hat U^\mathrm{ad}_\mathcal{S}(t,t')\left[ \mathds{1} + \hat V(t,t') \right]+\mathcal{O}\left(\epsilon^2/g^4\right)\label{Eq:US},
\end{equation}
where $\hat U^\mathrm{ad}_\mathcal{S}(t,t')$ is the evolution operator in leading order,
\begin{equation}
    \hat U^\mathrm{ad}_\mathcal{S}(t,t') =\sum_{a=\pm} {\vert a \rangle_{t} \, {}_{t'}\langle a \vert e^{-i\mu_a(t,t')}},
\end{equation}
and is thus diagonal in the instantaneous eigenbasis. The global phase factor $\mu_a(t,t')$ is the sum of the dynamic and geometric components,
\begin{equation}
    \mu_a(t,t') = \int_{t'}^{t}{\mathrm{d}\tau \left[ E_a(\tau) - i {}_{\tau}\langle a \vert \dot{a}  \rangle_{\tau} \right]}\,.
\end{equation}
Operator $\hat V(t,t')$ in Eq. \eqref{Eq:US} is of order $\epsilon/g^2$ and takes the form
\begin{equation}
    \hat V(t,t') = - {\alpha_{+-}(t,t')\vert + \rangle_{t'}\, {}_{t'}\langle -\vert}-{\rm H.c.}\,,
\end{equation}
with the coefficient $\alpha_{+-}(t,t')$ \cite{de_grandi_adiabatic_2010}
\begin{equation}
    \alpha_{+-}(t,t') =\dfrac{1}{2}\int_{t'}^{t}{\mathrm{d}\tau \dfrac{g\epsilon}{g^2 + \epsilon^2\tau^2}\exp\left[i\int_{t'}^\tau{\mathrm{d}u \sqrt{g^2 + \epsilon^2 u^2}} \right],\notag }
\end{equation}
and $\alpha_{-+}(t,t')=-\alpha^*_{+-}(t,t')$. In this procedure we account for non-adiabatic effects during the relaxation time of the meter by treating them to first order in the perturbative expansion. See App.\ \ref{App_B} for further details.

\subsubsection{Master equation}

The resulting master equation for the qubit density matrix in the laboratory frame and for $\tau_M\to 0$ takes the form
\begin{equation}
\label{Eq:NAME}
\partial_t\hat\rho=\mathcal L^{\rm ad}(t)\hat\rho\,,
\end{equation}
with generator
\begin{equation}
\mathcal L^{\rm ad}\hat{\rho} = -i \left[ \hat{H}_S, \hat{\rho} \right] -\gamma(t)\left( \hat{P}_-(t) \hat{\rho} \hat{P}_+(t) + \hat{P}_+ (t)\hat{\rho} \hat{P}_-(t) \right)\,,\label{Eq:dephasing}
\end{equation}
with operators $\hat{P}_{\pm}(t) = \vert \pm \rangle_{t} \, {}_{t}\langle \pm \vert$ projecting into the instantaneous basis with the time-dependent rate $$\gamma (t) = G(0)(E_+(t) - E_-(t))^2/2\,.$$
The scaling factor $G(0)/2$ is the real part of the Fourier transform $\Gamma_{XX}(\omega=0)$ of the meter's autocorrelation function at zero-frequency,
\begin{equation}
\Gamma_{XX}(\omega)=\int_0^{+\infty}{\mathrm{d}\tau \exp{(i\omega\tau)} \mathcal{
C}_{XX}(\tau,0)} \,.
\end{equation}
Note that $G(0)$ is here assumed to be positive, namely, the bath correlation functions are of positive-type, so that $\gamma(t)>0$. The quantum adiabatic master equation, Eq. \eqref{Eq:dephasing} describes an effective dephasing with time-varying rate $\gamma(t)$ in the instantaneous eigenbasis of $H_\mathcal{S}(t)$. The dephasing rate decreases with the instantaneous gap and is minimum at the anticrossing point $t=0$, where $\gamma(0)=G(0) g^2/2$. We use thereafter the notation $\gamma_0\equiv\gamma(0)$ for quantifying the strength of dephasing.

We note that, for the specific model of Eq. \eqref{Lindblad}
\begin{equation}
    G(0)= x_0^2(2n+1)\dfrac{\kappa}{(\kappa/2)^2 + \omega_c^2}\label{Eq_gamma}\,,
\end{equation}
where the details of the derivation are reported in App. \ref{App_C}. The dephasing rate monotonously increases with $x_0^2$ and with the thermal occupation $n$. This term results in an imaginary component of the gap between the instantaneous eigenstates, analogously to the dynamics discussed in Ref. \cite{Militello:2019}, whose amplitude increases with the temperature. As we will argue below, large reservoir temperatures result in lower infidelity, and thus larger adiabatic transfer probability. This regime corresponds to regions of Fig.\ \ref{fig:5} where the NM measure is the smallest, namely for $\kappa/\omega_c >0.3$ and $x_0 <0.5$.

\subsubsection{Adiabatic transfer in the Markovian regime}

\begin{figure}
\centering
\includegraphics[width = 0.95\columnwidth]{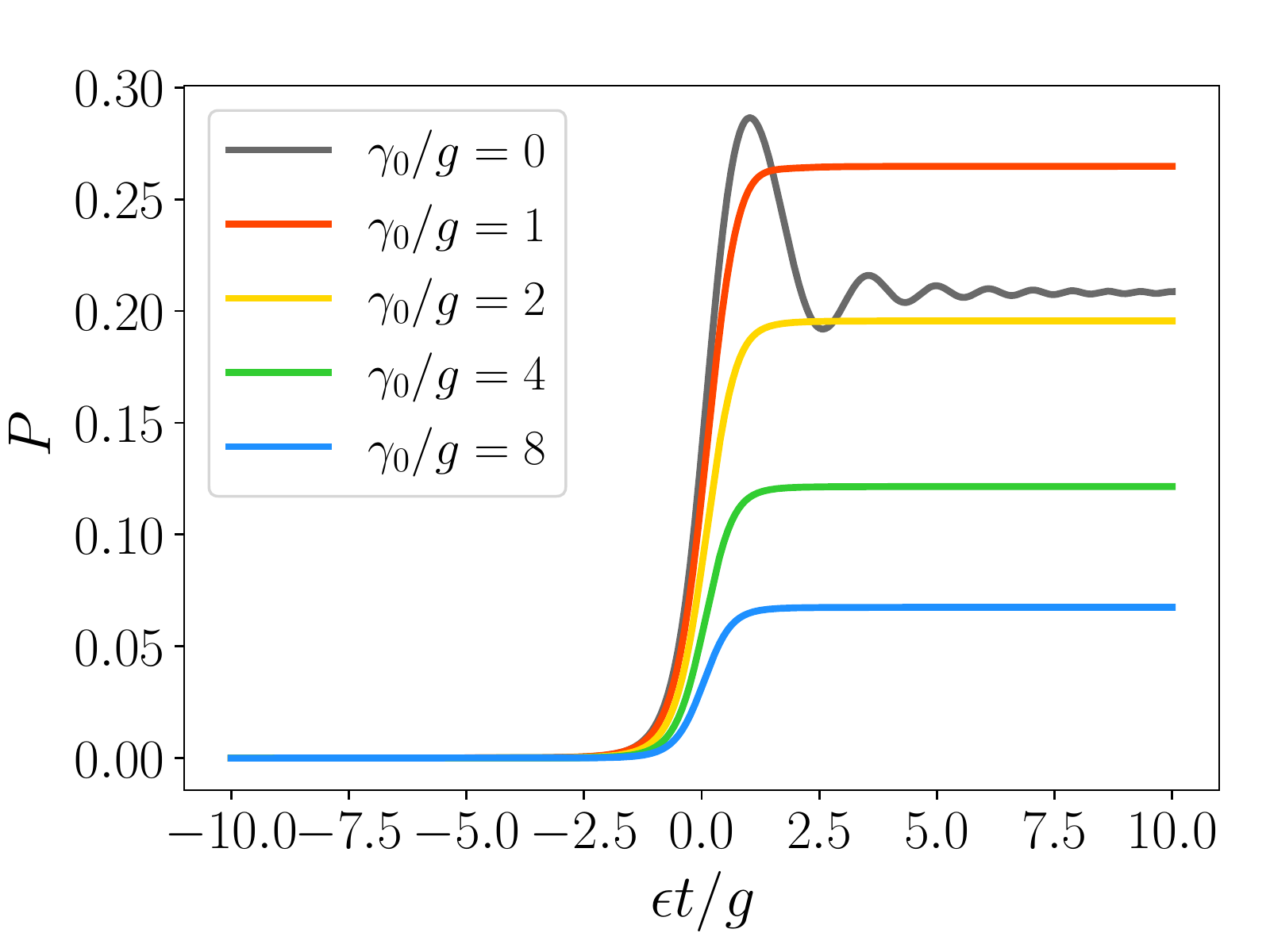}
\caption{(color online) Influence of the dephasing Lindbladian, Eq. \eqref{Eq:dephasing} on the evolution of the transition probability $P$ for a finite time evolution and different values of the ratio  $\gamma_0/g$ between the dephasing rate and the LZ coupling $g$. The curves are obtained by integrating numerically Eq.~\eqref{Eq:dephasing} for the initial state $|-\rangle_{t_i}$ over the finite time-window $[t_i,t_f]$ with $t_f=-t_i=5 g/\epsilon$ and $g^2/\epsilon = 1$. In all numerical calculations here and later on $\epsilon = 1$.}
\label{Fig4}
\end{figure}
The master equation \eqref{Eq:dephasing} was analysed in Ref. \cite{avron_landau-zener_2011,Novelli:2015}, with the noticeable difference that the dephasing rate was taken to be constant. In  Ref. \cite{avron_landau-zener_2011}  the asymptotic behaviour $T$ of the infidelity was analytically determined when $\sqrt{\epsilon}\ll g, \gamma_0$ (see also Refs. \cite{Avron2012, Fraas2014, Joye, Fraas2017} for related work). In this limit the infidelity can be cast into the form
\begin{equation}
T = \dfrac{\epsilon}{2g^2}Q\left( \dfrac{\gamma_0}{g}\right) + \mathcal{O}\left(\frac{\epsilon^2}{g^2(\gamma_0^2 +g^2)}\frac{\gamma_0^2}{\gamma_0^2 +g^2}\right)\label{Eq_6}\,,
\end{equation}
where $Q$ is an analytic function \cite{avron_landau-zener_2011}:
\begin{equation*}
    Q(x) = \dfrac{\pi}{2}\dfrac{x\left( 2 + \sqrt{1 + x^2} \right)}{\sqrt{1 + x^2}\left(\sqrt{1 + x^2} + 1 \right)^2}\,.
\end{equation*}
At fixed sweeping $\epsilon$, the behaviour of the infidelity $T$ is then controlled by a competition between the gap $g$ and the dephasing rate $\gamma_0$. One can especially notice that the first correction to Eq.~\eqref{Eq_6} exhibits a scaling with $\epsilon^2$ that echoes the one observed for the transition probability for a Landau-Zener model evolved on a finite time window with a linear ramp,  as in Eq. \eqref{Eq:finite} \cite{de_grandi_adiabatic_2010}. In particular, when the dephasing rate $\gamma_0 \gg g$, then to leading order $T \simeq \pi \epsilon/(4\gamma_0 g)$. This indicates an improvement of adiabaticity in the strong dephasing regime and is akin to the so-called quantum Zeno effect. Indeed, as each measurement projects the system onto an eigenstate of the measured observable, if the measurement is made at a fast frequency, the system does not have the time to evolve away from the state it was projected onto, thus suppressing the probability to tunnel toward another state.

We now analyse the behaviour of the infidelity beyond the regime of validity of Eq. \eqref{Eq_6} and study the competition between Hamiltonian dynamics and dephasing also for $\sqrt{\epsilon} \sim g, \gamma_0$. Figure~\ref{Fig4} displays the dynamics of the transition probability $P$ as a function of time for several ratios $\gamma_0/g$ and at fixed $\epsilon$. The effects of dephasing are two-fold: The oscillations of $P$ are damped down as $\gamma_0$ increases,  leading to a smoother dynamics. The steady state value is reached for transfer times of the order of $t_f\sim g/\epsilon$. Furthermore, the final value of $P$ decreases as the dephasing rate increases. For sufficiently large ratios $\gamma_0/g$ the infidelity $P$ is substantially reduced with respect to the value reached by the coherent Landau-Zener dynamics. 

\begin{figure}
\centering
\includegraphics[width = \columnwidth]{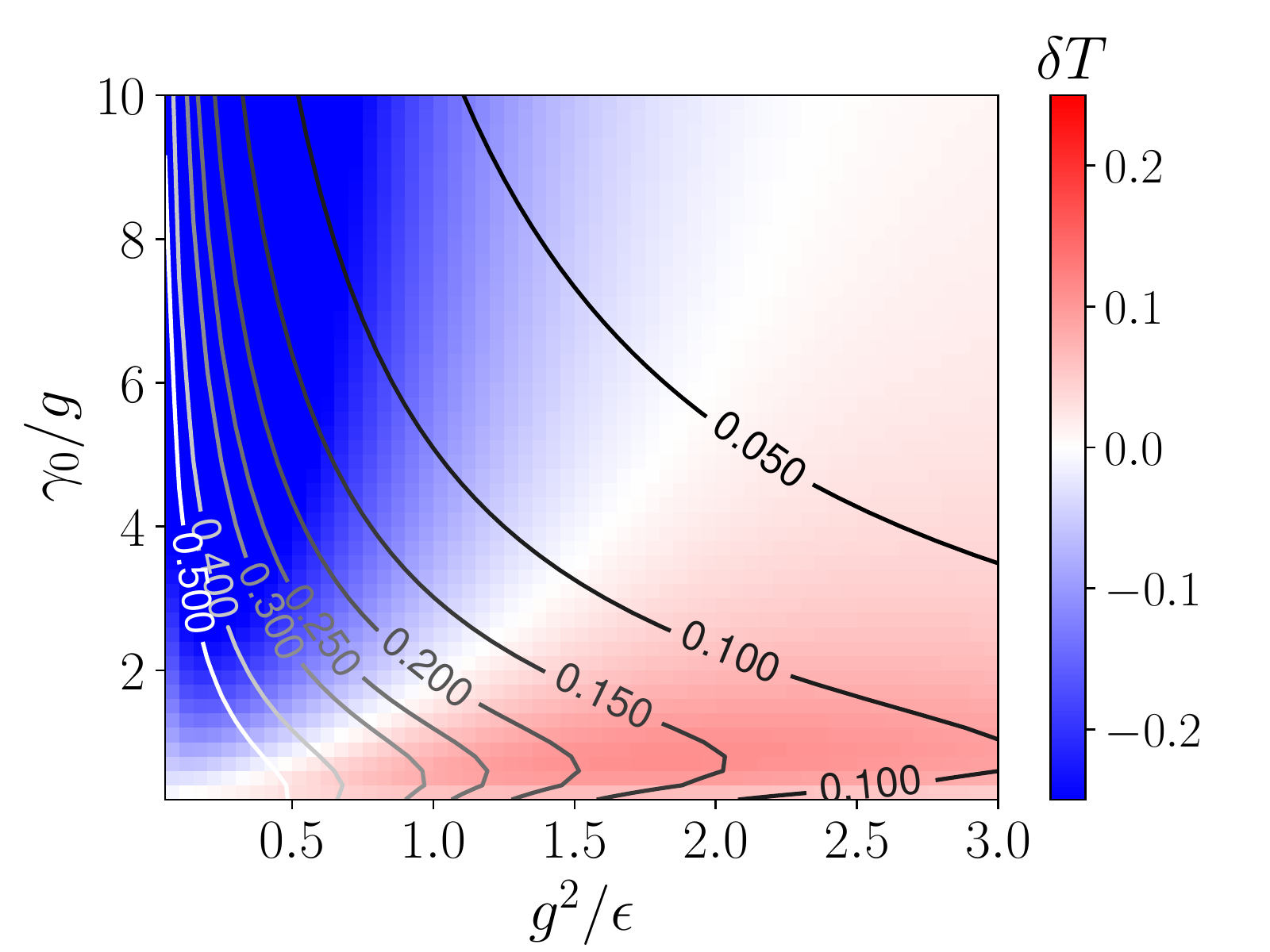}
\caption{(color online) Difference between the infidelity $T$ and the Landau-Zener prediction $T_{LZ}$, $\delta T=T - T_{LZ}$, as a function of $g^2/\epsilon$ and $\gamma_0/g$. The curves show the level lines of the infidelity $T$, the attached number reports the corresponding value. The infidelity $T$ is determined by numerically solving Eq. \eqref{Eq:dephasing} over the time interval  $[t_i,t_f]$ with $t_f=-t_i=5 g/\epsilon$. The transition probability $T_{LZ}$ used to compute $\delta T$ is the transfer probability for the coherent dynamics at the corresponding parameters. Note that the time interval over which the system is evolved is proportional to $g/\epsilon$.}
\label{Fig2}
\end{figure}

We analyse the relative infidelity $\delta T = T - T_{LZ}$ resulting from integrating Eq. \eqref{Eq:dephasing}. A negative value of this quantity indicates that the effect of the dephasing favors the adiabatic transfer with respect to the coherent Landau-Zener dynamics. The relative infidelity is displayed in Fig.~\ref{Fig2} as a function of the effective dephasing rate, $\gamma_0/g$ and of the adiabaticity parameter, $g^2/\epsilon$. The transfer time is here half of the one in Fig.~\ref{Fig4}. The behaviour at $\gamma_0=0$ is the prediction $T_{LZ}$ of the coherent Landau-Zener dynamics, where the infidelity decreases as $\epsilon$ decreases. By adding dephasing, for any value of $\epsilon$ the fidelity as a function of $\gamma_0$ first becomes worse ($\delta T >0$), then improves ($\delta T <0$). The difference $\delta T$ highlights the parameter region where the QND coupling improves the protocol's fidelity at finite times: we observe $\delta T <0$ for relatively large values of $\epsilon$, thus for relatively fast drives. In this regime dephasing  suppresses tunneling to the higher energy state,  as visible from the level lines of the infidelity in Fig.~\ref{Fig2}. For instance, an infidelity of $0.05$ is found even for $g^2/\epsilon<1$ by tuning the dephasing rate $\gamma_0$ to values $\gamma_0>10g$.

Via the derivation of an adiabatic master equation, we showed that the adiabaticity of the Landau-Zener dynamics is enforced by dephasing effects in the instantaneous LZ basis of the qubit. This dephasing mechanism is interpreted in terms of the Zeno effect and leads to a fast convergence of the transition probability $P$ towards its asymptotic value.

\section{Experimental implementations}
\label{Sec:Exp}

The dynamics discussed so far requires the capability to continuously tune the coupling between meter and qubit as a function of time. The experimental realization of a such a measurement protocol would require a continuous and perfect control over the parameters of the system over time, which is a challenging task. In this section we analyse the efficiency of the protocol when the coupling with the meter is implemented at certain instants of time. We first analyse the effect of the QND coupling as a function of the repetition rate of the measurement during the dynamics, assuming the capability to perfectly synchronize the measurement with the Landau-Zener evolution. We then investigate the infidelity when instead there is an error in the implementation, corresponding to an uncertainty in the exact time of the Landau-Zener evolution. 

\subsection{Stroboscopic QND measurement}

\begin{figure}
\centering
\includegraphics[width = \columnwidth]{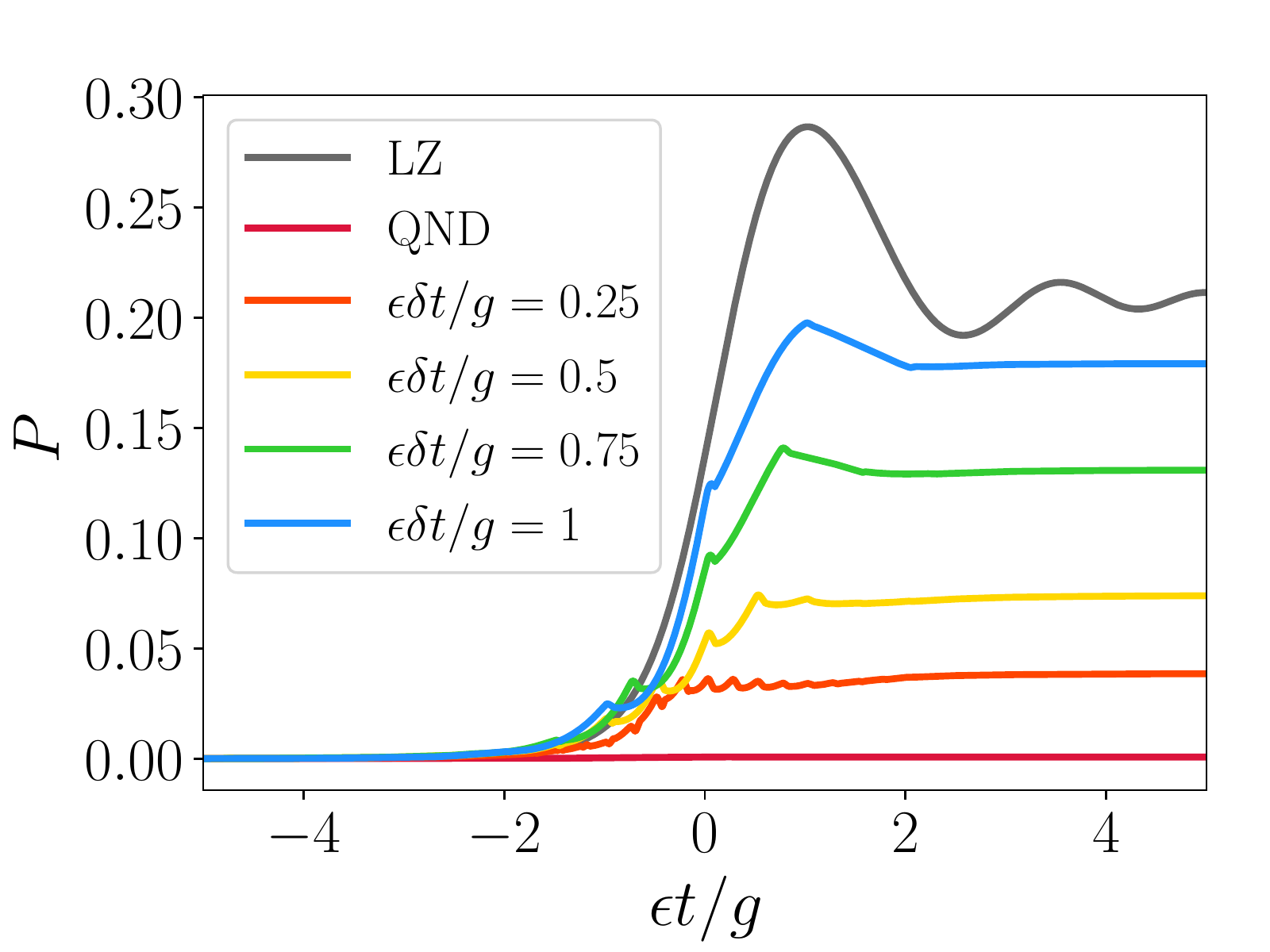}
\caption{Evolution of the transition probability $P$ for different sampling times $\delta t$. The results are compared with the Landau-Zener prediction (LZ) and the continuous QND measurement (QND). The parameters are $g^2/\epsilon = 1$, $\omega_c = g$, $\kappa = 2\omega_c$, $x_0 = 10$ and $n=0 $. The oscillator states are truncated at maximum occupancy $n_\mathrm{max} = 75$}
\label{Fig7}
\end{figure}

We now analyse the effect of a pulsed dynamics, such that the coupling with the meter is switched on at certain instants of time during the dynamics. We model the coupling using the Hamiltonian 
\begin{equation}
\hat H_{\rm QND}'(t)\simeq \sum_j \delta(t-j\delta t)x_0(\hat a +\hat a^\dagger)\otimes\hat H_{\mathcal S}(t)\,,
\end{equation}
where $\delta t$ is the time interval between two successive pulses. The effect of this discretized measurement is investigated numerically, approximating the Dirac distribution $\delta$ by a short pulse of duration $T_P= 1/x_0$. The outcome of these calculations is displayed in Fig.~\ref{Fig7}. The transition probability $P$ as a function of time exhibits cusps at the corresponding QND pulse. Each of these pulses corrects the evolved state of the spin and suppresses the transition probability, with an efficiency that increases as $1/\delta t$ decreases. Note that few QND measurements in the anticrossing region tend to increase the fidelity of the process and suppress the LZ oscillations. Even for sparse measurements the infidelity can lie below the value predicted for the bare Landau-Zener model.

\begin{figure*}
\centering
\includegraphics[width = 0.9\textwidth]{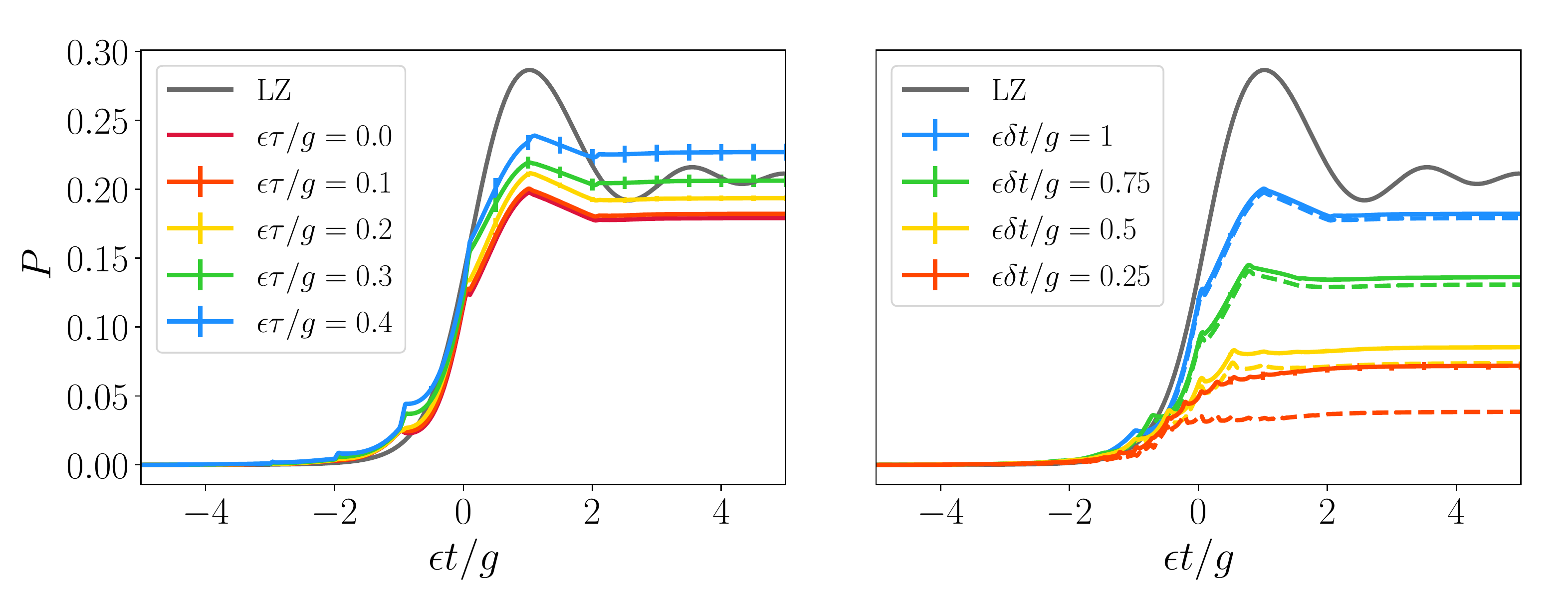}
\caption{Evolution of the transition probability $P$ at $g^2/\epsilon = 1$ for (a) different standard deviations $\tau$ for a fixed sampling time $\epsilon \delta t /g = 1$ and (b) for different sampling times $\delta t$ at fixed standard deviation $\epsilon \tau/g = 0.1$. The dashed lines correspond to the results for a perfect stroboscopic measurement. All curves are computed with the parameters as in Fig.\ \ref{Fig7}, and averaged over $n_\mathrm{it}=50$ samples of time-shifts. The curves also include the error bars on the sample-averaged values of $P$. In some cases, the errors are such that the bars merge with the plot line.}
\label{Fig8}
\end{figure*}

\subsection{Measurement errors}

We now consider the possibility of errors in the pulsed QND measurement. The error is modelled by a random time-shift in the qubit-meter coupling Hamiltonian, such that the QND Hamiltonian is not synchronized with the time evolution of the Landau-Zener Hamiltonian,
\begin{equation}
    \hat H_{\rm QND}''(t)\simeq \sum_j \delta(t-j\delta t)x_0(\hat a +\hat a^\dagger)\otimes\hat H_{\mathcal S}(t + t_j),
\end{equation}
where $\lbrace t_j \rbrace$ is a set of random variables uniformly distributed over the interval $[-\sqrt{3}\tau, +\sqrt{3}\tau]$ and $\tau$ is the standard deviation of the probability distribution. Figure~\ref{Fig8} shows the evolution of the transition probability $P$ for $n_\mathrm{it}=50$ samples. As visible in  Fig. \ref{Fig8}(a), for a fixed sampling frequency (here $\epsilon\delta/g t=1$), the introduction of errors leads to only a moderate increase of the infidelity, highlighting the robustness of the stroboscopic measurement against errors. Figure \ref{Fig8} (b) compares the evolution of the transition probability for several sampling frequencies at a fixed standard deviation ($\epsilon \tau/g = 0.1$) with the case of a perfect measurement. One can observe that the deviation from the ideal case is minimal for large sampling frequencies and grows with the sampling frequency.
The asymptotic behaviour of the infidelity remains comparable with the ideal case as long as the standard deviation of the time-shift is well below the sampling time $\tau \ll \delta t$. This demonstrates the robustness of the protocol against moderate parameter fluctuations.

\section{Conclusion and outlook}
\label{Sec:Conclusions}

We have proposed a protocol that can enforce adiabaticity in Landau-Zener dynamics by a quantum-non-demolition type of setup. The QND coupling we considered is time dependent and commutes with the Landau-Zener Hamiltonian at the given time. In the limit in which the meter instantaneously relaxes to its steady state, the QND coupling realizes an effective dephasing of the qubit in the instantaneous, adiabatic basis, thus suppressing diabatic transitions for given transfer times. In this regime, the transfer fidelity at finite times increases with the temperature of the thermal bath, with which the meter equilibrates. This result is consistent with studies on thermally assisted quantum annealing \cite{Amin2008, Amin2009}.

Interesting dynamics are found in the regime where the meter's relaxation time cannot be neglected over the characteristic time scale of the qubit. By suitably choosing the parameters, these dynamics perform an effective error correction by cooling the qubit to the lower instantaneous eigenstate, thus realising high fidelity adiabatic transfer in relatively short times. Previous work on Landau-Zener dynamics in the presence of external baths identified the competition of incoherent processes, which promote or suppress diabatic transitions \cite{Shimshoni1991,shimshoni_dephasing_1993} and cast them in terms of interference processes by the means of an elegant path-integral formulation. Our study shows that the fidelity of the transfer can be partially understood in terms of an effective gap induced by the time-dependent QND coupling. Future work will focus on analysing the connection between this gap and the spectral gap. One intriguing perspective is to be able to design an effective gap for achieving high-fidelity adiabatic transfer by means of an external environment.

The dynamics studied here is a realization of control of a quantum system by means of driven-dissipative dynamics in the spirit of Refs. \cite{Vacanti:2014,Alipour:2020}. It could be implemented in several setups, such as a single trapped ion \cite{Leibfried:2003}, a single trapped atom inside a resonator \cite{haroche2013exploring,Reiserer:2015}, and a superconducting qubit in circuit QED \cite{Atia:2020}. The QND type of Hamiltonian discussed here can extend the protocol of Ref. \cite{yang_quantum_2020} to tune the coupling between meter and qubit as a function of time. Errors in realising the stationary QND Hamiltonian have been discussed in Ref. \cite{yang_quantum_2020}. In the case here discussed they are systematically corrected by cooling generated by retardation effects in the coupling with the meter.

In view of practical applications, we have further shown that the requirement of continuous, time-dependent QND coupling can be relaxed: diabatic transitions can be suppressed by performing a stroboscopic series of instantaneous QND measurements during the dynamics. We have also included the effect of the fluctuations in implementing the specific form of the QND coupling and shown that the protocol is robust against timing errors.

The adiabatic transfer can be further optimized by tailoring the temporal variation of the Landau-Zener Hamiltonian \cite{PhysRevA.65.042308, Morita:2007, tian_universal_2018}, and combining measurements with optimal control techniques \cite{Horn_2018}. Suppression of errors at faster tuning rates can be studied beyond adiabatic perturbation theory in the framework of quantum non-adiabatic master equations \cite{PhysRevA.98.052129, PhysRevResearch.3.013064,Dupays:2020}. Future work will extend this protocol for reservoir engineering fast adiabatic dynamics across gapless points in many-body quantum systems.

\section*{Acknowledgments}

We thank Jacek Dziarmaga, Rosario Fazio, Ronnie Kosloff, and Marek Rems for their helpful insights. This work was funded by the Deutsche Forschungsgemeinschaft (DFG, German Research Foundation), Project-ID 429529648 TRR 306 QuCoLiMa ("Quantum Cooperativity of Light and Matter"), TRR 183 "Entangled States of Matter", and by the German Ministry of Education and Research (BMBF) via the Projects NiQ ("Noise in Quantum Algorithms") and Quantera "NAQUAS". Project NAQUAS has received funding from the QuantERA ERA-NET Cofund in QuantumTechnologies implemented within the European Union's Horizon 2020 Programme. 

\appendix

\section{The Born-Markov approximation}
\label{App_A}
In the following, we will derive the Lindblad equation in the case of a system coupled to a measurement apparatus which performs non-demolition measurement.

The construction of the non-adiabatic master equation requires some assumptions on the different energy and time-scales involved in the description of dynamics, starting with the minimal gap $g$ between the two energy branches. The gap shall be compared with the rate at which the Hamiltonian $H_\mathcal{S}$ evolves, namely $\epsilon$. In order to acertain the adiabaticity of the coherent part of the dynamics for the LZ system, we shall make sure that $\epsilon/g^2 \ll 1$. Under this condition, the incoherent part of the dynamics is dominated by adiabatic mechanisms since non-adiabatic effects scale like $\epsilon/g^2$. The derivation of the adiabatic master equation will also requires that the meter relaxes at time scales $\tau_M$ over which one can consider the Hamiltonian $H_\mathcal{S}(t)$ to be constant, resulting in $\epsilon\tau_M^2\ll 1$.

Let us consider $\hat{\chi}$, the density matrix representing the two-level system and the measurement apparatus, such that the density matrix of the system is obtained via the partial trace of the states of the meter $\hat{\rho}(t) = \mathrm{Tr}_{\mathcal{M }}[\hat{\chi}(t)]$. The time-evolution of the density matrix $\hat{\chi}$ is piloted by the Liouville-von Neumann equation
\begin{equation}
    \dfrac{\mathrm{d}}{\mathrm{d} t}\hat{\chi} = -i\left[\hat{H}_{\mathcal{SM}}, \hat{\chi}\right],
\end{equation}
where $H_{\mathcal{SM}}$ is a Hamiltonian of the generic form
\begin{equation}
       \hat{H}_{\mathcal{SM}} = \hat{H}_\mathcal{S}(t) \otimes \hat{\mathds{1}}_\mathcal{M} + \hat{\mathds{1}}_\mathcal{S} \otimes \hat{H}_\mathcal{M} + \hat{H}_\mathcal{S}(t) \otimes \hat{X}_\mathcal{M}.
\end{equation}
Let us notice that we are actually treating an unusual case where the Hamiltonian of the system is time-dependent, and so is the interaction term that couples the system to the meter, via non-demolition measurement.

\bigskip
To treat the effects of this coupling on the dynamical properties of the system, we will place ourselves in the interaction picture, thus isolating the interaction dynamics from the free dynamics. In the interaction picture, we may define the family of freely evolved operators $\tilde{A}(t)$ such that
\begin{equation}
 \tilde{A}(t) = \hat{U}^\dagger(t,t_0)\hat{A}(t_0)\hat{U}(t,t_0),
\end{equation}
where the free evolution operator $U(t,t_0)$ is defined as the solution of the Schr\"odinger equation in the non-interacting case
\begin{align}
    \hat{U}(t,t_0) &= \mathcal{T}\exp\left[-i\int_{t_0}^t{\mathrm{d}t'\left(\hat{H}_{\mathcal{S}}(t') + \hat{H}_{\mathcal{M}}\right)}\right]\\
    &= \hat{U}_\mathcal{S}(t,t_0)\otimes \hat{U}_\mathcal{M}(t,t_0)\notag,
\end{align}
where $\mathcal{T}$ is the time-ordering operator.

\bigskip
The Liouville-von Neumann equation in the interaction pictures is then modified, as terms describing the coherent part of dynamics are eliminated by the application of the free evolution operator, only leaving terms accounting for the incoherent dynamics. The equation of motion of the density matrix can then be self-consistently expanded up to an arbitrary order
\begin{align}
    \dfrac{\mathrm{d}}{\mathrm{d} t}\tilde{\chi}(t) 
    &= -i\left[\tilde{V}_{\mathcal{SM}}, \tilde{\chi}(0)\right]\notag\\ 
    &- \int_0^t{\mathrm{d}t' \left[ \tilde{V}_{\mathcal{SM}}(t), \left[\tilde{V}_{\mathcal{SM}}(t'), \tilde{\chi}(t')\right]\right]}
\end{align}
with $\tilde{V}_{\mathcal{SM}} =  \tilde{H}_\mathcal{S} \otimes \tilde{X}_\mathcal{M} $ being the interaction between the system and the measurement device. Under the assumption that the system is initially decoupled to the meter, tracing out the meter will result on the vanishing of the first term in the previous equation. The equation of motion for the reduced density matrix then reads
\begin{equation}
    \dot{\tilde{\rho}}(t) = - \int_0^t{\mathrm{d}t' \mathrm{Tr}_{\mathcal{M}}\left\lbrace\left[ \tilde{V}_{\mathcal{SM}}(t), \left[\tilde{V}_{\mathcal{SM}}(t'), \tilde{\chi}(t')\right]\right]\right\rbrace}.
\end{equation}
In order to proceed forward, we will need to make a first assumption on the form taken by the full density matrix $\tilde{\chi}$.

\bigskip
\underline{\textit{Hypothesis 1:}} Assuming that the coupling between the system and the meter is sufficiently weak in order to leave the state of the meter globally constant over time, we may write the density matrix as the product state of the system and the meter
\begin{equation}
    \partial_t\tilde{\chi(t)} = \tilde{\rho}(t)\otimes \hat{M}_0 + \mathcal{O}\left(\tilde{V}_{\mathcal{SM}}\right).
\end{equation}
Under this assumption, tracing out the meter leads to a description of the incoherent dynamics in terms of the time-correlation properties of the meter. Developing the two commutators, one obtains the following master equation for the evolution of the reduced density matrix 
\begin{align}
    \partial_t\tilde{\rho}(t) = - \int^t_0{\mathrm{d}t'}& \left\lbrace \tilde{H}_\mathcal{S} (t) \tilde{H}_\mathcal{S} (t')\tilde{\rho}(t')\right. \\
    &\left. -\tilde{H}_\mathcal{S} (t') \tilde{\rho}(t') \tilde{H}_\mathcal{S} (t)\right\rbrace \mathcal{C}_{XX}(t,t')\notag \\
    -\int^t_0{\mathrm{d}t'}&\left\lbrace \tilde{\rho}(t')\tilde{H}_\mathcal{S} (t') \tilde{H}_\mathcal{S} (t)\right.\\
    &\left. -\tilde{H}_\mathcal{S} (t) \tilde{\rho}(t') \tilde{H}_\mathcal{S} (t')\right\rbrace \mathcal{C}_{XX}(t',t),\label{EqA_8}
\end{align}
where $\mathcal{C}_{XX}(t,t')= \langle \tilde{X}_\mathcal{M}(t) \tilde{X}_\mathcal{M}(t') \rangle$ is the auto-correlation function of the observable $X_\mathcal{M}$ that couple the meter to the system.

\bigskip

In order to accurately approximate the dynamics of the open system $\mathcal{S}$ by a Markovian process, one has to assume the time-scale separation of the system and the meter, or in other words that the dynamics of the meter is much faster than the one of the system. Therefore, the information transferred from the system to the meter is quickly erased and has no backward effect on the dynamics of the system, then resulting in the loss of memory that characterizes a Markovian time-evolution.

\bigskip

\underline{\textit{Hypothesis 2:}} Assuming that the dynamics of the meter is much faster than the one of the system, the correlations of the meter are expected to decay sufficiently fast in time so that we may approximately treat the density matrix as a constant over the integration time:
\begin{equation}
    \tilde{\rho}(t') \simeq \tilde{\rho}(t)
\end{equation}
 in the integral term of Eq.~\eqref{EqA_8} whose upper bound can be taken to infinity. After a change of variable ($t' = t - \tau$) and the application of the time-scales separation approximation, the quantum master equation now takes the form given in Eq. \eqref{Eq:MEq}.

Usually, assuming the weak-coupling of the system and its environment and time-scale separation is sufficient in order to transform this integro-differential equation into a Lindblad equation. However, in order to perform the integrals we have to make further assumptions.

\section{Adiabatic evolution approximation}
\label{App_B}

In order to give the quantum master equation the form of a Lindblad equation, we will take advantage of the slow nature of the dynamics of the system. Indeed, assuming that the evolution of the system is adiabatic, we may provide an approximate expression for the evolution operator $U_\mathcal{S}(t,t')$, thus simplifying the calculations.

\bigskip
\underline{\textit{Hypothesis 3:}}  At first order in the adiabatic expansion, the evolution operator reads
\begin{equation}
    \hat{U}_\mathcal{S}(t,t')=\hat{U}^\mathrm{ad}_\mathcal{S}(t,t')\left[ \hat{\mathds{1}} + \hat{V}(t,t') \right]\label{Eq_15}\,,
\end{equation}
where the contribution at leading order reads
\begin{equation}
    \hat{U}^\mathrm{ad}_\mathcal{S} =\sum_{a} {\vert a \rangle_{t} \, {}_{t'} \langle a \vert e^{i\mu_a(t,t')}}\,,
\end{equation}
and the correcting term $\hat{V}(t,t')$ takes the form 
\begin{equation}
\label{eq:V}
    \hat{V}(t,t') = - \sum_{a \neq b}{\alpha_{ab}(t,t')\vert a \rangle_{t'} \, {}_{t'} \langle b\vert}\,.
\end{equation}
Both operators are here expressed in the basis of the  instantaneous eigenstates of the Hamiltonian $\hat{H}_\mathcal{S}(t)$, such that $\hat{H}_\mathcal{S}(t)\vert E_a(t)\rangle = E_a(t)\vert E_a(t) \rangle$. The scalar $\mu_a(t,t')$ is here the phase accumulated in the time interval $[t',t]$:
\begin{equation}
    \mu_a(t,t') = \int_{t'}^{t}{\mathrm{d}\tau \left[ E_a(\tau) - i {}_{\tau}\langle a \vert \dot{a} \rangle_{\tau} \right]}.
\end{equation}
and the coefficients $\alpha_{ab}(t,t')$ in Eq. \eqref{eq:V} are expressed as
\begin{equation}
    \alpha_{ab}(t,t') = \int_{t'}^{t}{\mathrm{d}\tau e^{-i(\mu_b(\tau,t') - \mu_a(\tau,t')} {}_{\tau}\langle a \vert \dot{b}\rangle_{\tau} }.
\end{equation}

Before applying this approximate expression in the study of the quantum master equation, we will use the properties of the evolution operator to formulate one last approximation. Indeed, the following reasoning, we will be brought to encounter $\hat{U}_\mathcal{S}(t-\tau,0) = \hat{U}_\mathcal{S}(t-\tau,t)\hat{U}_\mathcal{S}(t,0)$. Yet, since the meter is assumed to evolve much faster than the system, we allow ourselves to assume that the Hamiltonian stays constant over the time necessary for the correlations $\mathcal{C}_{XX}$ to decay, we may then state that
\begin{equation}
    \hat{U}_\mathcal{S}(t-\tau,0) \simeq e^{i\tau \hat{H}_\mathcal{S}(t)}\hat{U}^\mathrm{ad}_\mathcal{S}(t,0).
\end{equation}

\bigskip
In what follows we detail how the adiabatic approximation modifies the first of the integral terms and then generalize it to all the right-hand side of Eq.~\eqref{Eq:MEq}. The first integral terms then becomes
\begin{widetext}
\begin{align}
    &\int_0^{+\infty}{\mathrm{d}\tau \hat{U}^{\mathrm{ad} \dagger}_\mathcal{S}(t,0) e^{-i\tau \hat{H}_\mathcal{S}(t)} \hat{H}_\mathcal{S}(t)e^{i\tau \hat{H}_\mathcal{S}(t)}\hat{U}^\mathrm{ad}_\mathcal{S}(t,0)}\tilde{\rho}(t)\tilde{H}_\mathcal{S}(t)\mathcal{C}_{XX}(\tau,0)\notag\\
    &\simeq \sum_{a,b} e^{i\mu_{ab}(t,0)} \vert a \rangle_{t} (H_\mathcal{S}(t))_{ab} \, {}_{t}\langle b\vert\tilde{\rho}(t)\tilde{H}_\mathcal{S}(t)\int_0^{+\infty}{\mathrm{d}\tau e^{i\tau\left[ E_b(t) - E_a(t)  \right]} \mathcal{C}_{XX}(\tau,0)}\notag\\
    &\simeq \sum_a {E_a(t) \vert a \rangle_{t} \, {}_{t} \langle a\vert \tilde{\rho}(t) \tilde{H}_\mathcal{S}(t)\int_0^{+\infty}{\mathrm{d}\tau \mathcal{C}_{XX}(\tau)}}\notag\\
     &\simeq \sum_{a,b} {\Gamma_{XX}(0) E_a(t) E_b(t) \vert a \rangle_{t=0} \, {}_{t=0} \langle a \vert \tilde{\rho}(t) \vert b \rangle_{t=0} \, {}_{t=0}\langle b \vert }\notag,
\end{align}
\end{widetext}
where $\mu_{ab}(t,0) = \mu_a(t,0) - \mu_b(t,0)$ and the matrix element $(H_\mathcal{S}(t))_{ab} = \langle a(t)\vert H_\mathcal{S} \vert b(t) \rangle = E_a(t) \delta_{ab}$. Furthermore, we note the spectral function of the auto-correlation $\Gamma_{XX}(\omega)=\int_0^{+\infty}{\mathrm{d}\tau \exp{(i\omega\tau)} \mathcal{
C}_{XX}(\tau,0)}$. Following the same path for the other integral terms, the quantum master equation then takes the compact form
\begin{equation}
 \partial_t\tilde{\rho} = \sum_{a,b}{\Gamma_{XX}(0)E_a(t)E_b(t) \hat{P}_a(t)\left[\tilde{\rho}(t), \hat{P}_b(t) \right]} + \mathrm{h.c.}\label{Eq_A16},
\end{equation}
where $\hat{P}_a(t) = \vert a(t) \rangle \langle a(t) \vert$ is the projector at time $t$ onto the eigenstate labeled $a$.

\bigskip
From this point, we can return to the Schr\"odinger picture and recast this quantum master equation into the form of a Lindblad equation describing dephasing mechanisms. Using the relationship $\tilde{\rho}(t) = \hat{U}^\dagger_\mathcal{S}(t,0)\hat{\rho}(t) \hat{U}_\mathcal{S}(t,0)$, the left-hand side of Eq.~\eqref{Eq_A16} transforms back into the Liouville-von Neumann part of the Lindblad equation
\begin{equation}
    \hat{U}_\mathcal{S}(t,0) \partial_t\tilde{\rho}\hat{U}^\dagger_\mathcal{S}(t,0) = \partial_t\hat{\rho} + i\left[ \hat{H}_\mathcal{S}(t), \hat{\rho}(t) \right].
\end{equation}
The application of the evolution operator to the right-hand side of Eq.~\eqref{Eq_A16} on the other hand results in bringing the projectors $P_a$ to time $t$
\begin{align}
     \partial_t\hat{\rho} &= - i\left[ \hat{H}_\mathcal{S}(t), \hat{\rho}(t) \right] \notag \\
     &+\sum_{a,b}{\Gamma_{XX}(0)E_a(t)E_b(t) \hat{P}_a(t)\left[\rho(t), \hat{P}_b(t) \right]} + \mathrm{h.c.},
\end{align}
where the spectral function $\Gamma_{XX}(0)$ can be split into two different contributions $ \Gamma_{XX}(0) = \dfrac{1}{2}G(0) + i S(0)$.
both defined from the full Fourier transform of the auto-correlation function
\begin{subequations}
\begin{align}
    G(\omega) & = \int_{-\infty}^{+\infty}{\mathrm{d}\tau e^{i\omega\tau}\mathcal{C}_{XX}(\tau, 0)}\\
    S(\omega) &= \int_{-\infty}^{+\infty}{\dfrac{\mathrm{d}\omega'}{2\pi}G(\omega')\mathcal{P}\left( \dfrac{1}{\omega - \omega' }\right) },
\end{align}
\end{subequations}
where $\mathcal{P}$ is Cauchy principal value.

In the case of the Landau-Zener model, the spectrum is reduced to only two levels $E_{\pm}(t)$, such that $E_+(t) = - E_-(t)$. The quantum master equation then takes the form of Eq. \eqref{Eq:dephasing}.

%

\section{Lindblad equation in the adiabatic limit}
\label{App_C}
\begin{figure}
    \centering
    \includegraphics[width=\columnwidth]{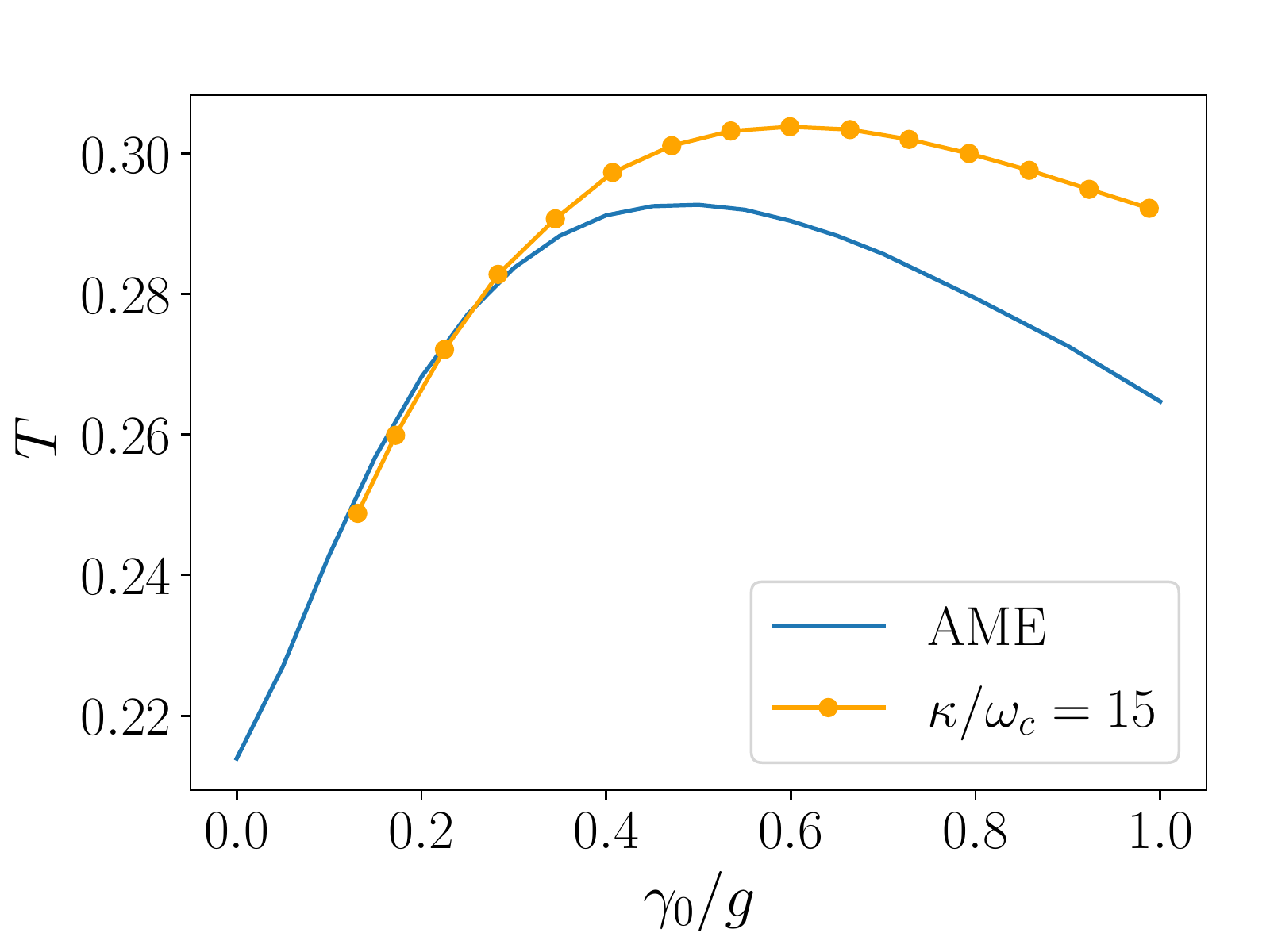}
    \caption{Asymptotic infidelity $T$ as a function of the dephasing rate $\gamma_0$. Fixing the values of $g^2/\epsilon = 1$, $x_0 = 1$, $\omega_c = g$ and $\kappa$, the dephasing rate at anticrossing $\gamma_0$ is tuned by changing the temperature, which results in a shift of the vibrational occupancy $n$.}
    \label{FigApp1}
\end{figure}

We will discuss in the following the Lindblad equation in its adiabatic regime, namely in the limit when the damping rate $\kappa$ is large, our purpose being to acertain the fact that dephasing effects become then predominant and that the behaviour of the Lindblad equation is consistent with the predictions of the adiabatic master equation for the qubit. The correlation function associated with the observable $\hat{X} = x_0(\hat{a} + \hat{a}^\dagger)$ can be computed in the case of the damped harmonic oscillator described by the Lindblad equation
\begin{align}
    \partial_t \hat{\rho} = &-i\left[ \omega_c \hat{a}^\dagger \hat{a}, \hat{\rho} \right] + \kappa(n+1)\left( \hat{a}\hat{\rho} \hat{a}^\dagger - \dfrac{1}{2}\lbrace \hat{a}^\dagger \hat{a}, \hat{\rho} \rbrace \right) \notag \notag \\
    & +\kappa n \left( \hat{a}^\dagger\hat{\rho} \hat{a} - \dfrac{1}{2}\lbrace \hat{a} \hat{a}^\dagger, \hat{\rho} \rbrace \right)\label{EqC1}.
\end{align}
The derivation of the correlation function relies on the decomposition of the density matrix in the right and left eigenvectors of the Liouvillian superoperator $\rho_\lambda$ and $\check{\rho}_\lambda$, respectively, such that $\mathcal{L}\rho_\lambda = \lambda \rho_\lambda$ and $\check{\rho}_\lambda \mathcal{L} = \lambda \check{\rho}_\lambda$, with ${\rm Tr}\{\check{\rho}_\lambda{\rho}_\lambda\}=\delta_{\lambda,\lambda'}$. This decomposition is known for Eq. \eqref{EqC1} \cite{Briegel:1993,Englert:2003}.

The correlation function that we defined as $\mathcal{C}_{XX}(t,t')=\langle X(t) X(t') \rangle$ is expressed in terms of averages over the equilibrium state of Eq.~\eqref{EqC1}, namely the thermal state $R_0=\exp(-\beta\omega_c a^\dagger a)/Z$. The non-vanishing contributions of the correlation function then lead to the form
\begin{align}
   \mathcal{C}_{XX}(t,t') = & \langle x_0^2(\hat{a}(t) + \hat{a}^\dagger(t))(\hat{a}(t') + \hat{a}^\dagger(t'))\rangle\notag\\
   = & x_0^2\left( \mathrm{Tr}\lbrace \hat{a}^\dagger(t) \hat{a}(t')  \hat{R}_0 \rbrace + \mathrm{Tr}\lbrace \hat{a}(t) \hat{a}^\dagger(t')  \hat{R}_0 \rbrace  \right)\notag\,.
\end{align}
Using the completeness relation of the eigenbasis $\sum_\lambda \rho_\lambda \check{\rho}_\lambda = \mathds{1}$, we decompose the two contributions of the correlation function as
\begin{equation*}
    \mathrm{Tr}\lbrace \hat{a}^\dagger(t) \hat{a}(t')  \hat{R}_0 \rbrace = \sum_{\lambda}\mathrm{Tr}\lbrace \hat{a}^\dagger e^{(t-t')\mathcal{L}}  \rho_\lambda \rbrace \mathrm{Tr}\lbrace \check{\rho}_\lambda \hat{a}  \hat{R}_0 \rbrace,
\end{equation*}
where trace is performed over the basis of the coherent states. Due to orthogonal properties of the Laguerre polynomial involved in the eigenoperators of the Liouvillian, all the contributions in the sum vanish except for the one corresponding to $(k=\pm 1, n=0)$, leading to the simple expression for the correlation function
\begin{equation*}
    \mathcal{C}_{XX}(t,t')=x_0^2 \,e^{-\kappa(t-t')/2} \left((n+1)e^{-i\omega_c(t-t')} + n e^{i\omega_c(t-t')} \right)\,.
\end{equation*}
As a result, we obtain that the real part of the correlation function at frequency $\omega=0$ reads
\begin{equation}
    \dfrac{G(0)}{2} = x_0^2(2n+1)\dfrac{\kappa/2}{(\kappa/2)^2 + \omega_c^2}\label{Eq_gamma:1}\,,
\end{equation}
from which can be deduced the value of $\gamma_0$, the dephasing rate at the anticrossing point. The asymptotic value of the infidelity $T$ computed via the adiabatic master equation and the Lindblad equation is displayed on Fig.~\ref{FigApp1} and shows a consistent behaviour of infidelity in both cases: $T$ increases for weak values of the dephasing rate before reaching a maximum and decreasing.

\bibliography{main}

\end{document}